# Phonon Spectra and Vibrational Heat Capacity of Quasi-One-Dimensional Structures Formed by Rare Gas Atoms on the Surface of Carbon Nanotube Bundles

E.V. Manzhelii, I.A.Gospodarev, S.B. Feodosyev

*B. Verkin Institute for Low Temperature Physics and Engineering of the National Academy of Sciences of Ukraine, 47 Nauky Ave., Kharkiv 61103, Ukraine*

*E-mail:*emanzhelii@ilt.kharkov.ua

**Abstract**

The features of phonon spectra and their effect on the vibrational heat capacity of linear chains of inert gas atoms adsorbed onto a substrate, which is the surface of nanotubes bound to a nanobundle. The influence of the substrate results both in a shift of the lower limit of the chain spectrum from zero, and in mechanical stress in the chain (its extension or compression) also. It is shown that in the case of a compressed chain, the non-central interaction between atoms is negative (repulsive), it results in a shift of the lower boundary of the spectrum of transverse vibrations to low frequencies and to a shortening of the part of the specific heat temperature dependence in which this dependence is close to exponential. Heterogeneity of the nanobundle structure can cause a change in the distances between atoms of the chain. It is shown both and analytically and numerically, that as a result of it, discrete levels with frequencies both above and below the quasi-continuous spectrum band can appear in the phonon spectrum of the chain. The discrete levels with frequencies below the quasi-continuous spectrum band lead to a further shortening of the temperature interval at which the temperature dependence of the specific heat is close to the exponential one.

# Introduction

Quasi-one-dimensional (q1D) crystalline structures attract great interest from both the fundamental and applied points of view, in particular, as promising materials for quantum computers. The interest is caused by the unique properties of their quasiparticle spectra, such as the root singularities of the spectral densities at the edges of quasicontinuous spectrum bands, the thresholdless formation of discrete levels localized near defects, etc. Due to the Landau-Peierls instability [1], the existence of these structures is impossible without some three-dimensional substrate, the choice of which is associated with considerable difficulties. The substrate should ensure the stability of q1D systems of sufficient length and, at the same time, minimally distort its 1D spectral peculiarities.

Recently the adsorption of rare-gas atoms onto carbon nanotube bundles is often used to obtain stable macroscopically long q1D structures [2-8]. At low concentration of adsorbed gases, their atoms are located in grooves between nanotubes on the nanobundle surface (and in the case of 4He, inside nanobundles and even inside nanotubes). In the grooves between the nanotubes, the adsorbed atoms can form linear chains of length ~ 10 μm. The length corresponds to the number of atoms in the chain ~ $10^3$-$10^4$. For the chains of this length, the boundary effects can be neglected. The one-dimensional nature of these structures is confirmed by both neutron-diffraction studies [2] and heat capacity data [3-7]. Neutron diffraction studies of $^4$He atoms adsorbed in grooves on the nanobundle surface have shown the periodicity of the arrangement of $^4$He atoms in the chain [9]. Theoretical calculations have shown the presence of a periodic potential along the grooves on the surface of nanobundles [8]. The variation of amplitude of this potential depends on the relative orientation and displacement of nanotubes forming the groove. The potential depth varies from the values slightly greater than zero to 40K. All this makes it possible to describe the vibrational characteristics of the adsorbed chains within the harmonic dynamics of the crystal lattice.

It was shown in [10] that starting with a certain frequency $\omega_0$, the vibrations of the linear chain deposited on the crystal surface or in the bulk, actually do not extend through the crystal matrix and are completely localized on the chain. The frequency $\omega_0$ is determined by the contribution of the interaction of an atom in the chain with the atoms of the crystal-matrix in the self-interaction matrix of the atom in the chain, Thus, at $\omega > \omega_0$, neither the structure nor the phonon spectrum of the crystal-matrix can considerably change the spectral characteristics of atoms in the chain. The effect of a crystal-matrix on the phonon spectrum of the chain can be expressed in terms of only one parameter, namely, the value of the initial frequency $\omega_0$. This approximation is particularly profound for the vibrational characteristics of the chains of inert gas atoms adsorbed precisely on a carbon substrate because of the large difference between the Debye temperatures of inert gases and carbon structures.

At $\omega < \omega_0$, the vibrational spectrum of the atoms of the adsorbed linear chains has a three-dimensional character which determines the convergence of the mean-square displacements of atoms in the chain and the stability of these structures in a finite temperature range. The temperature dependence of the phonon heat capacity of the adsorbed linear chain necessarily contains a low-temperature interval in which the temperature dependence of the heat capacity is close to the exponential one.

At $\omega > \omega_0$, the vibrations of the atoms of the chain are either quasi-localized or their propagation has one-dimensional character and the spectral densities of these atoms are well described by simple analytical expressions obtained for one-dimensional models [11]. The bandwidth of the quasi-continuous spectrum of the chain is determined by the interaction between the atoms of the chain [10, 11], which depends on the distance $r$ between them. Naturally, this distance is significantly affected by the interaction of adsorbed atoms of the chain with the carbon atoms of nanobundles. Therefore, as a rule, $r$ does not coincide with the distance $r_0$ corresponding to the minimum of the interatomic interaction potential in the chain. At $r < r_0$, the parameter of noncentral interaction, which determines the width of the spectrum of transverse vibrations, is negative. It will result in a shift of the minimum frequency of the quasi - continuous spectrum to the low-frequency region [11] and, consequently, to the displacement of the linear part of the heat capacity [10] to the region of lower temperatures. [11]. The formation of a compressed chain ($r < r_0$) by atoms of inert gases seems quite plausible because the period of the field created by nanotubes in the grooves is smaller than the equilibrium distance for most inert gases [9]. Note that the negative parameter of the noncentral interaction is intrinsic to many solidified gases and metals [12, 13].

The defects of nanotubes, as well as the incommensurability of the periods of the adsorbed chain and of the field along the groove between the nanotubes, can cause a local change in the interaction of the atoms in the chain with nanotubes. In the case of a local change in this interaction, the distance between a pair of atoms in the chain can also change, which, in turn, can result in the appearing of a localized state with a frequency below the lower limit of the quasi-continuous spectrum of the chain [11]. It leads to a shift of the linear part of the heat capacity curve to the low-temperature region. The high-temperature part of the heat capacity curve of atomic chains adsorbed on nanobundles was studied in [14].

In this paper, we study the effect of the difference in the interatomic distance $r$ in the adsorbed chain from the distance $r_0$, corresponding to the minimum of the interaction potential between the atoms in the chain, on its phonon spectrum, formation conditions and characteristics of the discrete vibrational levels localized at the defects. The contribution of all these changes of the spectrum into the low-temperature heat capacity is also studied.

## Phonon spectrum and vibrational heat capacity of an ideal adsorbed atomic chain

It was shown in [10] that a chain of atoms adsorbed in a groove between nanotubes with a sufficiently high degree of accuracy can be considered as a chain in an external field that determines the initial frequency of its quasi-continuous spectrum. Basing on this result, we will consider the chain of atoms adsorbed in the grooves between nanotubes as a chain of atoms in an external periodic field. We take into account an interaction only between the nearest neighbors, which is quite natural for inert gases. In this case, the dispersion relations have the form:

$$\varepsilon_l(k) \equiv \omega_l^2(k) = \varepsilon_{0l} + \frac{4\beta_l}{m}\sin^2\frac{ka}{2};$$
$$\varepsilon_\tau(k) \equiv \omega_\tau^2(k) = \varepsilon_{0\tau} + \frac{4\beta_\tau}{m}\sin^2\frac{ka}{2}. \quad (1)$$

Here $\varepsilon \equiv \omega^2$ is the squared frequency, $k$ is the wave vector, the indices $l$ and $\tau$ correspond to the longitudinal and transverse vibrations respectively. For a pair-wise isotropic interaction between atoms, the parameter of the central interaction $\alpha$ and the parameter of the non-central interaction $\beta$ are expressed in terms of the potential of this interaction $\varphi(r)$ by the following range: $\beta_l(r) = \frac{\partial^2 \varphi(r)}{\partial r^2}$; $\beta_\tau(r) = \frac{1}{r}\frac{\partial \varphi(r)}{\partial r}$. We note that the symmetry condition for the tensor of elastic moduli should be applied to the whole system (including not only the chain, but also the substrate), since it is the interaction of the chain with the substrate that ensures the stability of the chain. Therefore, the transverse vibrations of the atoms in the chain (1) are not flexural vibrations, so their dispersion relation is not $\varepsilon_\tau(k) \sim k^4$ in the long-wavelength region.

Unlike the positive parameter of the central interaction $\beta_l$, the parameter of the non-central interaction $\beta_\tau$ changes its sign according to the relative position of the atom with respect to the minimum of the interatomic pair-wise potential of the interaction $\varphi(r)$. For the case of a compressed chain ($r < r_0$), the parameter of the non-central interaction is negative. Dispersion relations for $r > r_0$ and $r < r_0$ are presented in Fig.1. In what follows, the width of the continuous spectrum of longitudinal oscillations will be denoted by $\Delta_l$, and the width of the continuous spectrum of transverse oscillations, by $\Delta_\tau$ ($\Delta_l = 4\beta_l/m$, $\Delta_\tau = 4\beta_\tau/m$).

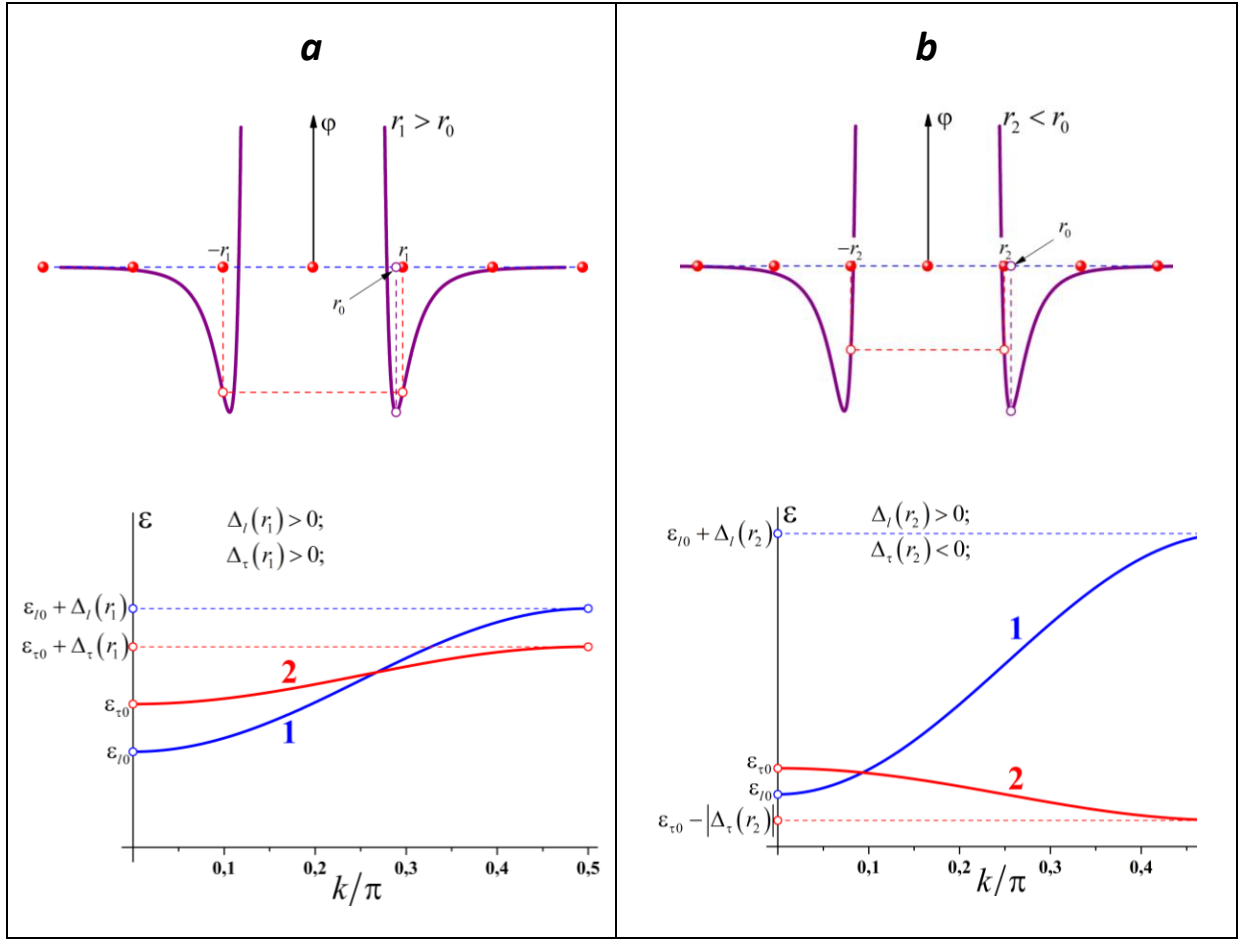

**Fig**. 1: The dispersion relations (1) and (2) are shown by the curves 1 and 2 in both fragments, for the cases $r > r_0$ (fragment a) and $r < r_0$ (fragment b), respectively.

In [11], the problem of the vibrational properties of both the ideal chain of atoms in an external field and a chain with a defect is solved by using the Jacobian matrix method [15-17]. The space of displacements of atoms of a chain is represented as a direct sum of orthogonal subspaces $H = H^{(-)} \oplus H^{(+)}$, where the subspace $H^{(-)}$ is the subspace of in-phase displacements of atoms, and the subspace $H^{(+)}$ is the subspace of anti-phase displacements. Each of them is a linear span of the sequence of vectors $\{\mathbf{L}^n h_0^{(-)}\}_{n=0}^{\infty}$ and $\{\mathbf{L}^n h_0^{(+)}\}_{n=0}^{\infty}$ respectively. Here, $h_0^{(-)}$ and $h_0^{(+)}$ are the so-called generating vectors, corresponding to in-phase and anti-phase displacements of two neighboring atoms, respectively, and $\mathbf{L}$ is the dynamic operator which has the form:

$$\mathbf{L}_{ik}(R,R') = \frac{(\varepsilon_{l0}+2\beta_l)\cdot\delta_{RR'} - \beta_l\left(\delta_{R,R'+a}+\delta_{R,R'-a}\right)}{m}\cdot\delta_{ix}\delta_{ik} + \\ + \frac{(\varepsilon_{l0}+2\beta_\tau)\cdot\delta_{RR'} - \beta_\tau\cdot\left(\delta_{R,R'+a}+\delta_{R,R'-a}\right)}{m}\cdot\delta_{iz}\delta_{ik}, \quad (2)$$

($R$ and $R'$ are the coordinates of the chain atoms).

Complete information about the vibrational spectrum of the chain is contained in its Green function $\hat{G}(\varepsilon) = (\varepsilon\hat{I}-\hat{L})^{-1}$, where $\hat{I}$ is the unit operator. In the formalism of the Jacobian matrices, all matrix elements of the operators $\hat{G}^{(\pm)}(\varepsilon)$ are expressed in terms of elements $G_{00}^{(\pm)}(\varepsilon) = \left(h_{i0}^{(\pm)}, \hat{G}(\varepsilon) h_{i0}^{(\pm)}\right)$ [18], for which we obtain [11]:

in the subspace $H^{(-)}$:

$$G_{i00}^{(-)}(\varepsilon) = \frac{2}{\Delta_i}\left[1 + Z_i(\varepsilon)\operatorname{sgn}(\varepsilon-\varepsilon_{i0})\sqrt{\left|\frac{\varepsilon-\varepsilon_{i\max}}{\varepsilon-\varepsilon_{i0}}\right|}\right]; \quad \Delta_i > 0;$$

$$G_{i00}^{(-)}(\varepsilon) = \frac{2}{|\Delta_i|}\left[-1 + Z_i(\varepsilon)\operatorname{sgn}(\varepsilon_{i0}-\varepsilon)\sqrt{\left|\frac{\varepsilon-\varepsilon_{i\min}}{\varepsilon_{i0}-\varepsilon}\right|}\right]; \quad \Delta_i < 0;$$

(3)

in the subspace $H^{(+)}$:

$$G_{i00}^{(+)}(\varepsilon) = \frac{2}{\Delta_i}\left[-1 - Z_i(\varepsilon)\operatorname{sgn}(\varepsilon-\varepsilon_{i\max})\sqrt{\left|\frac{\varepsilon-\varepsilon_{i0}}{\varepsilon-\varepsilon_{i\max}}\right|}\right]; \quad \Delta_i > 0;$$

$$G_{i00}^{(+)}(\varepsilon) = \frac{2}{|\Delta_i|}\left[1 + Z_i(\varepsilon)\operatorname{sgn}(\varepsilon-\varepsilon_{i\min})\sqrt{\left|\frac{\varepsilon-\varepsilon_{i0}}{\varepsilon-\varepsilon_{i\min}}\right|}\right]; \quad \Delta_i < 0.$$

(4)

Here $Z_i(\varepsilon) \equiv \Theta(\varepsilon_{i\min}-\varepsilon) + i\cdot\Theta(\varepsilon-\varepsilon_{i\min})\Theta(\varepsilon_{i\max}-\varepsilon) - \Theta(\varepsilon-\varepsilon_{i\max})$, and $\Theta(x)$ is the Heaviside $\Theta$-function. The values $\varepsilon_{i\min}$ and $\varepsilon_{i\max}$ are squares of the minimum and maximum vibration frequencies respectively. Figures 2 and 3 show the Green functions in the subspaces and for the longitudinal and transverse vibrations of the chain at $r > r_0$ and at $r < r_0$. We emphasize that for transverse vibrations $\Delta_\tau < 0$ and $\varepsilon_{\tau\min} = \varepsilon_{\tau 0} - |\Delta_\tau|$ for $r < r_0$.

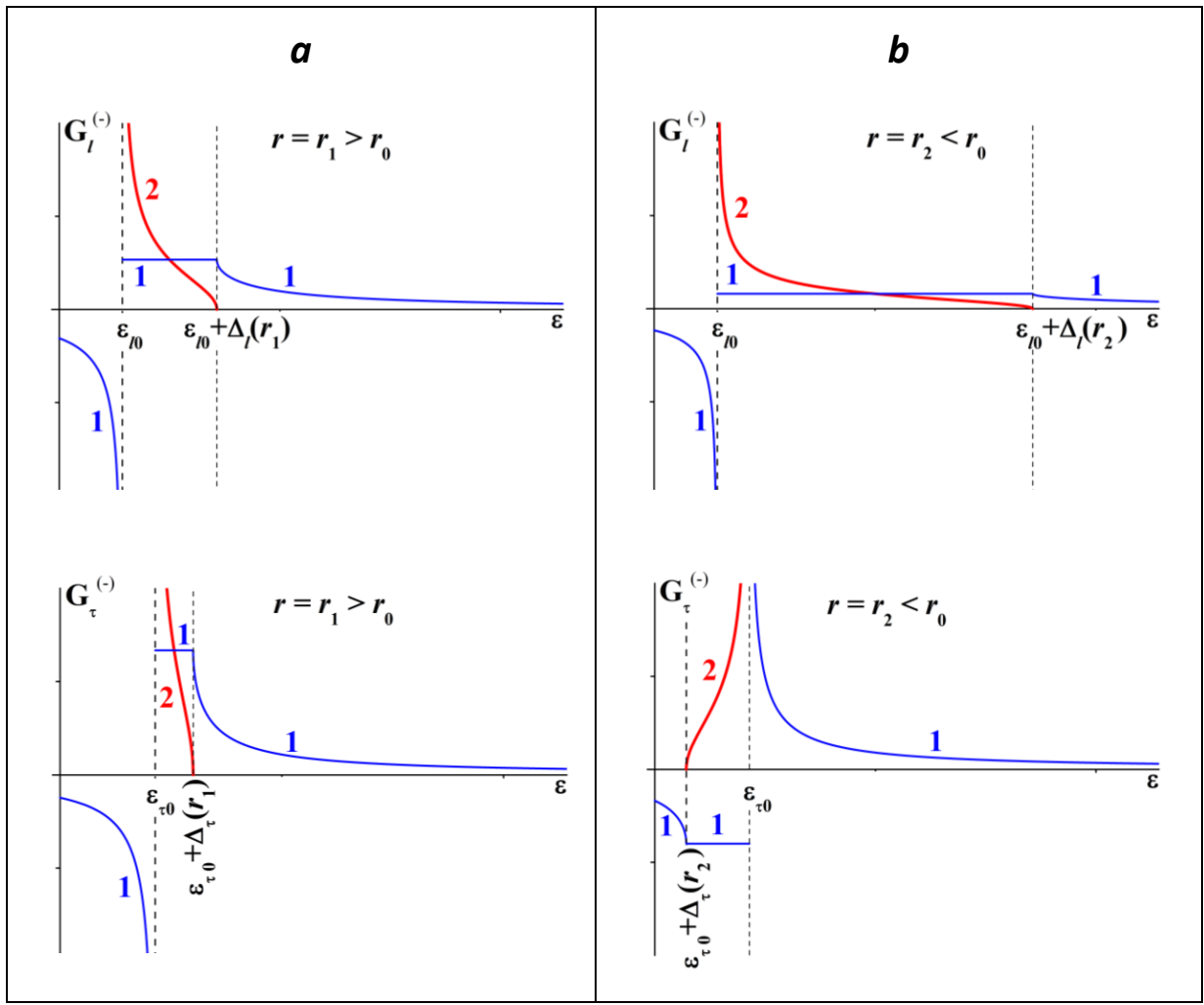

**Fig. 2:** Real (curves 1) and imaginary (curves 2) parts of Green's function (3) for the cases $r > r_0$ (fragment *a*) and $r < r_0$ (fragment *b*). The upper parts of the both fragments show the functions $G_l^{(-)}(\varepsilon)$; the lower parts show the functions $G_\tau^{(-)}(\varepsilon)$.

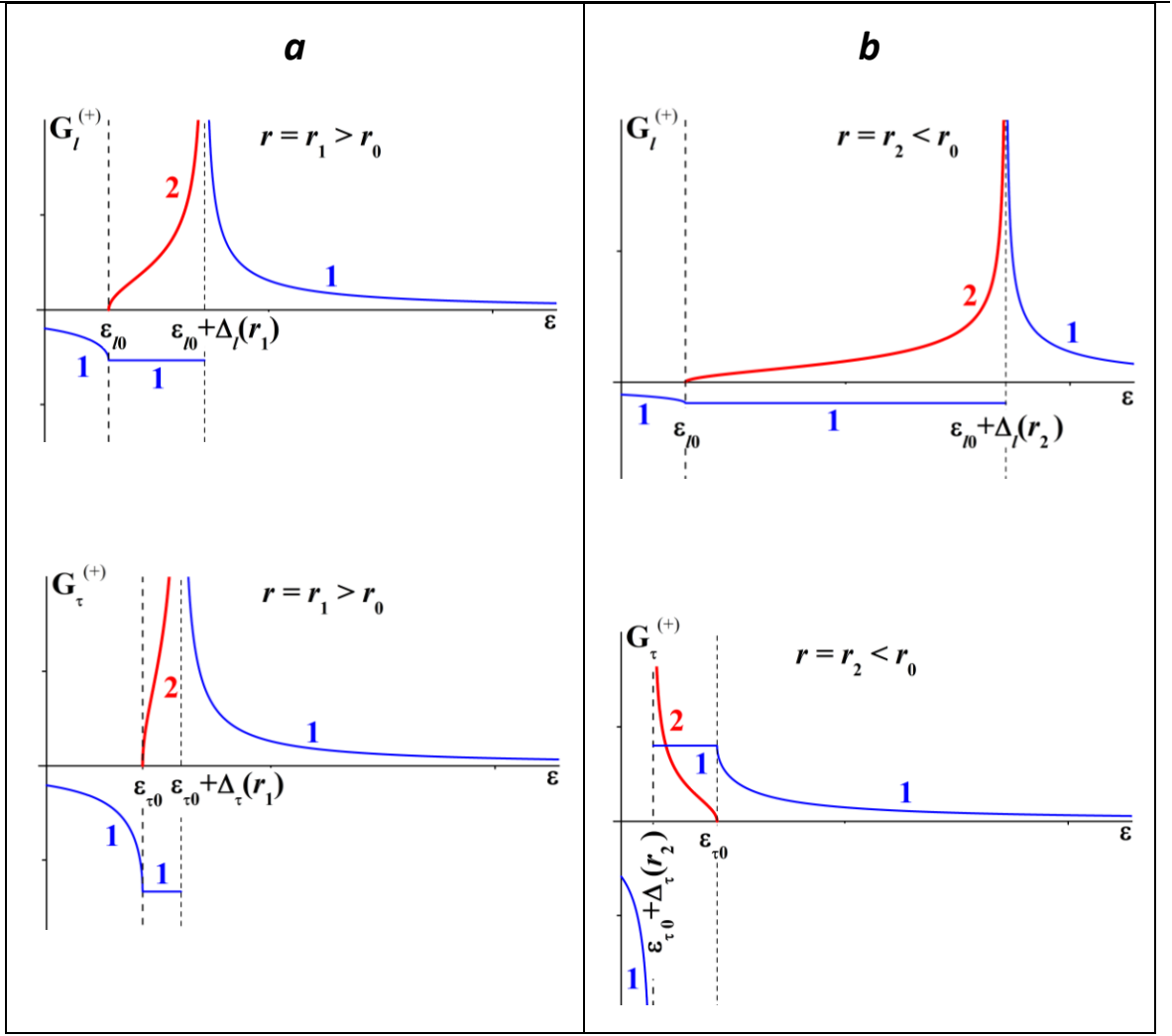

**Fig. 3:** Real (curves 1) and imaginary (curves 2) parts of Green's functions $G_i^{(+)}(\varepsilon)$ (4). Notation is completely analogous to the notation used in the previous figure.

The density of the vibrational states of the chain has the form:

$$g(\varepsilon) = \frac{1}{6\pi} \operatorname{Im}\left\{G_l^{(-)}(\varepsilon) + G_l^{(+)}(\varepsilon) + 2\left[G_\tau^{(-)}(\varepsilon) + G_\tau^{(+)}(\varepsilon)\right]\right\}.$$

Since this function contains the spectral densities of transverse and longitudinal vibrations, it has square root singularities not only at the edges of the spectrum, but also within it. Figure 4 shows the dependencies $g(\varepsilon)$ for $r > r_0$ and $r < r_0$. For $r < r_0$, the band of the quasi-continuous spectrum is much wider because of high values of the interaction parameter $\beta_l$, peculiar to these interatomic distances, which determines the longitudinal vibrations, and of negativity of the parameter $\beta_\tau$, which determines the transverse vibrations.

The spectral densities of oscillations shown in Fig. 4 correspond to heat capacity temperature dependences with different lengths of the parts that are close to exponential. Fig. 5 shows the temperature dependences of the contributions to the low temperature heat capacity of the chains $C_V(T) = C_{Vl}(T) + 2C_{V\tau}(T)$ of the longitudinal $C_{Vl}(T)$ and transverse vibrations $C_{V\tau}(T)$ (curves 1 and 2 respectively), and also $C_V(T)$-curves 3. Here $T$ is the temperature. The position of the linear-like part of $C_V(T)$ is determined by the position of the inflection point. When $r < r_0$, the initial frequency of the continuous spectrum shifts to low frequencies. It results in the fact that the linear-like part of the heat capacity curve starts at temperatures lower than for $r > r_0$. When $r < r_0$, the heat capacity curve $C_{Vl}(T)$ is flattened because of the large width of the quasi-continuous spectrum. At relatively high temperatures in both cases, the heat capacities $C_V(T)$ have close values.

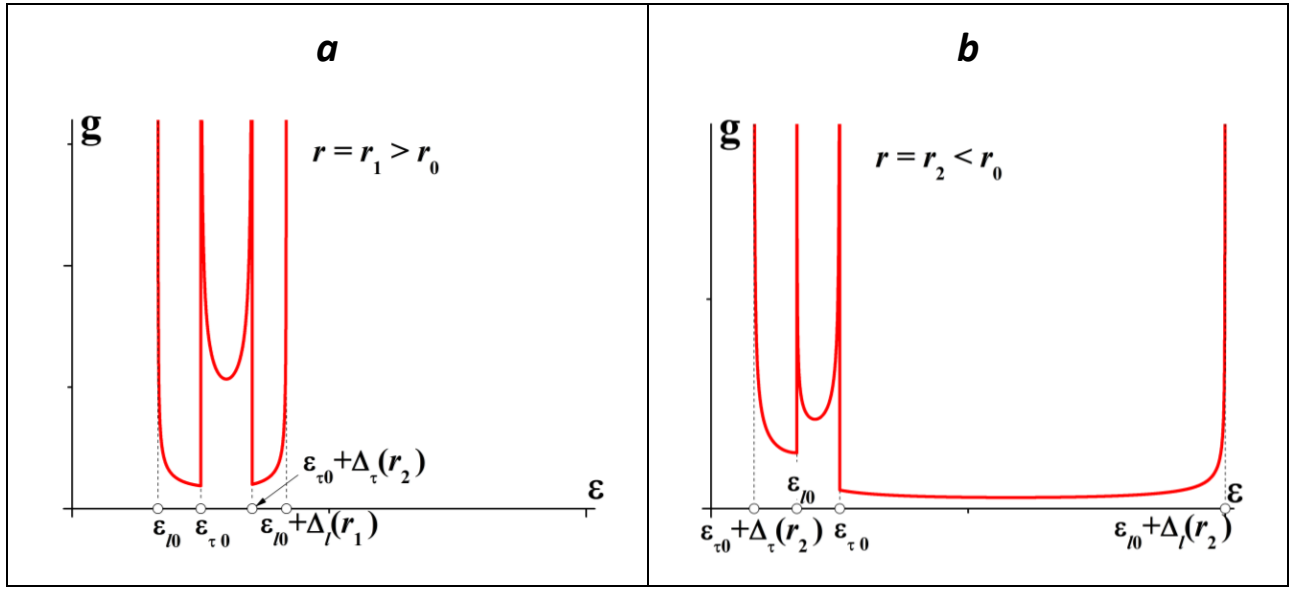

**Fig. 4:** Density of phonon states for $r > r_0$ (fragment *a*) and $r < r_0$ (fragment *b*).

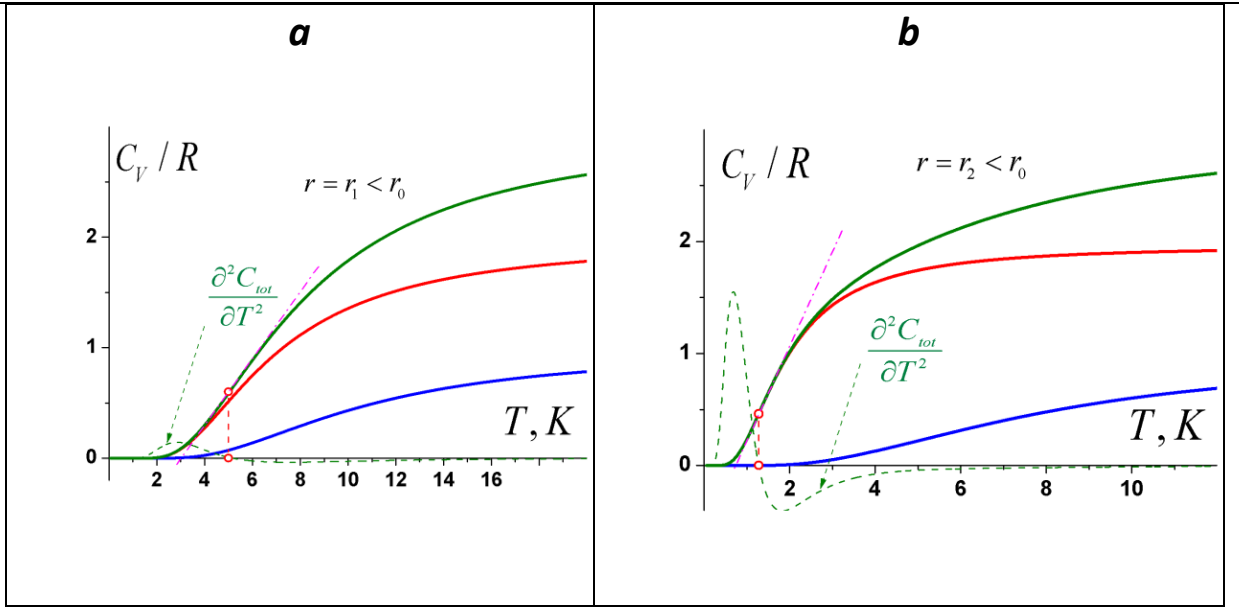

**Fig. 5:** Phonon heat capacity of the chain for $r > r_0$ (fragment *a*) and for $r < r_0$ (fragment b). In both fragments curves 1 are the contributions to the heat capacity of longitudinal vibrations; curves 2 are the contributions to the heat capacity of transverse vibrations; and curves 3 are the total heat capacities. The dashed line is the value proportional to $\dfrac{\partial^2 C_V(T)}{\partial T^2}$. The dash-dotted line is the tangent to $C_V(T)$ at the inflection point.

## The effect of a local change in the distance between a pair of atoms on the vibrational spectrum and the phonon heat capacity

Consider an adsorbed chain with a local defect, namely, an isolated pair of atoms the distance between which differs from the distance between the other neighboring atoms. In the technique of the Jacobian matrices this defect is a degenerate local perturbation [17, 19]. The frequencies of local vibrations caused by this defect can be obtained from the Lifshitz equation [20], which, applied to the problem in question, can be written as:

$$\operatorname{Re} G_i^{(\pm)}(\varepsilon) = 1/\lambda_i^{(\pm)}. \tag{5}$$

Here $\lambda_i^{(\pm)}$ are the operators describing the contribution of the defect to the operators $L_i^{(-)}$ and $L_i^{(+)}$. The operators $L_i^{(-)}$ and $L_i^{(+)}$ are induced by operator (2) for displacement along the crystallographic direction $i$ in the subspaces $H^{(-)}$ and $H^{(+)}$ respectively.

In the subspace of the in-phase vibrations $H^{(-)}$, the change in the interaction between the atoms of the chain defect is not shown. Therefore, the operator $\lambda_i^{(-)}$ is determined only by the change in the interaction of the chain with the substrate, that is, the change in the initial frequency of the quasi-continuous spectrum $\varepsilon_{i0\,def} = \varepsilon_{i0}(1+\delta_i)$. In this case, the operator $\lambda_i^{(-)} = \varepsilon_{i0}\delta_i$.

In the subspace $H^{(+)}$, both the change in the value of the interaction between the atoms of the defect with the substrate, and the change in the interaction between the atoms of the defect, is shown. This change can be written as $\Delta_{i\,def} = (1+\eta_i)\Delta_i$, and the operator $\lambda_i^{(+)}$ can be written as

$$\lambda_i^{(+)} = (\varepsilon_{i0}\delta_i + \frac{\Delta_i}{2}\eta_i) \ .$$

When studying the low-temperature heat capacity, the localized states lying below the initial frequency of the quasi-continuous spectrum are of considerable interest. For convenience, we will call them gap localized states. Denote the square of the frequency of the gap state by $\varepsilon_g$. The conditions for the existence of solutions of the Lifshits equation (8) can be easily obtained from the Green functions (5, 6) shown in Fig. 2.3. The detailed description of these conditions is given in [11]. In this paper, we dwell only on the conditions of their existence. Thus, because the value of $\operatorname{Re} G_i^{(-)}(\varepsilon_{\min})$ tends to $-\infty$ at $\Delta_i > 0$, the gap levels appear at any negative values of $\lambda_i^{(-)}$ (i.e., values of $\delta_i$). When $\Delta_i < 0$ (transverse vibrations at $r < r_0$), the gap vibrations localized on the in-phase vibrations of atoms of the chain appear at $\delta_i < -\frac{\Delta_i}{2\varepsilon_{i0}}$. For anti-phase vibrations, the gap levels appear at any negative values of $\lambda_i^{(+)}$ at $\Delta_\tau < 0$. In other cases, in the subspace $H^{(+)}$ gap vibrations occur at $\delta_i < -\frac{(1+\eta_i)\Delta_i}{2\varepsilon_{i0}}$.

The energy values of the gap vibrational levels are the poles of the Green function of the perturbed system $\tilde{G}_{i00}^{(\pm)}(\varepsilon, \lambda_i^{(\pm)}) = \left(h_0^{(\pm)}, \left[\varepsilon I - L_i^{(\pm)} - \lambda_i^{(\pm)}\right]^{-1} h_0^{(\pm)}\right)$. The residues at these poles are the so-called intensities $\mu_{d0}$ of these levels. If the local spectral density in a defect-free chain $\rho_{i00}^{(\pm)}(\varepsilon) = \pi^{-1} \operatorname{Im} G_{i00}^{(\pm)}(\varepsilon)$ is normalized to unit, then the relation $\int_D \tilde{\rho}_{i00}^{(\pm)}(\varepsilon) d\varepsilon = 1 - \mu_{d0}$ is valid for the local spectral density of the defect $\tilde{\rho}_{i00}^{(\pm)}(\varepsilon)$. The positive values of intensities $\mu_{id} > 0$ correspond to the existence of discrete vibrational levels. Hence, the energy values of the gap levels are:

$$\varepsilon_{ig}^{(\pm)} = \left( a_i^{(\pm)} + \frac{\left(\Lambda_i^{(\pm)}\right)^2 + b_i^2}{\Lambda_i^{(\pm)}} \right), \quad \Lambda_i^{(\pm)} < 0.$$

The condition for the existence of the gap levels ($\mu_{id} > 0$) has the form:

$$\Lambda_i < -|b_i|.$$

It was also shown in [11] that in each of the subspaces, one phonon splits off from the quasi-continuous spectrum band to the gap level.

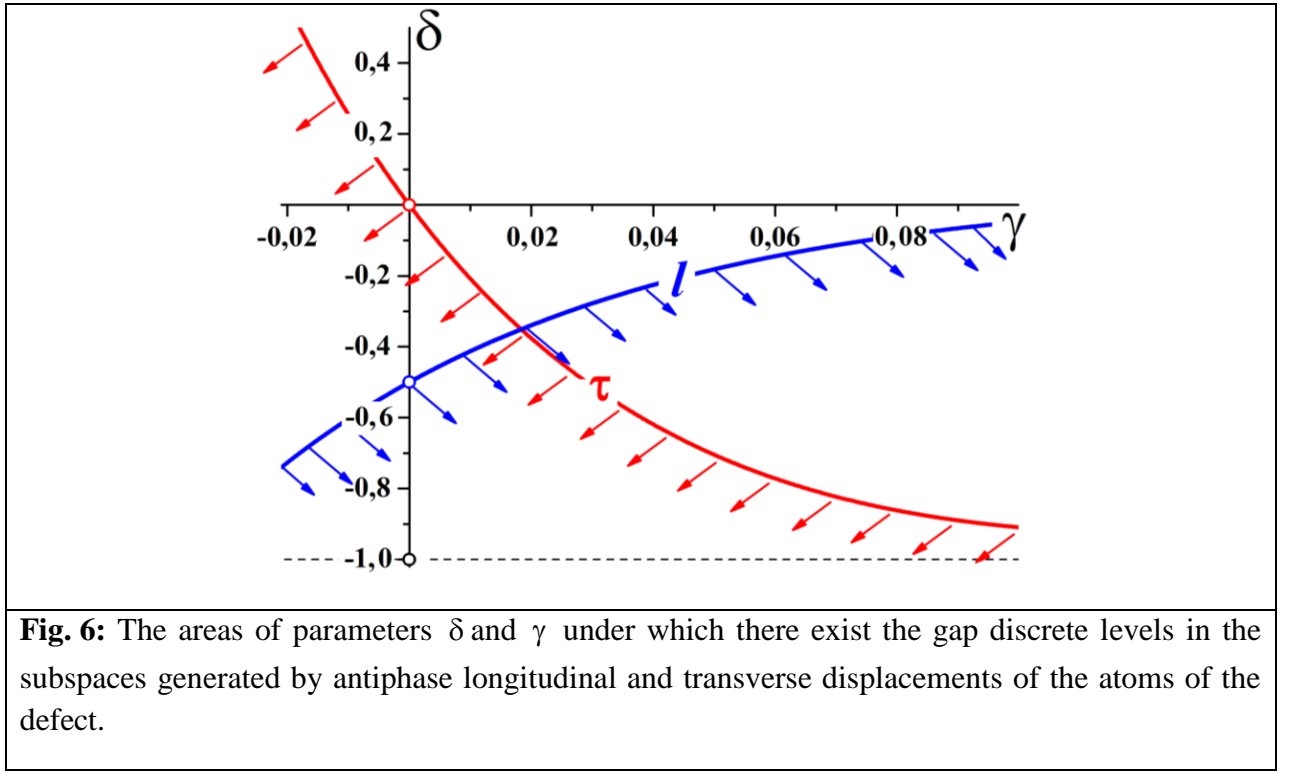

**Fig. 6:** The areas of parameters $\delta$ and $\gamma$ under which there exist the gap discrete levels in the subspaces generated by antiphase longitudinal and transverse displacements of the atoms of the defect.

From the Lennard-Jones potential and the condition for the existence of gap levels, one can obtain a diagram of the existence of gap levels localized in anti-phase vibrations of the chain (Fig.6) Vibrations localized on defects cause a change in thermodynamic quantities and, in particular, in heat capacity. The change in the phonon spectrum due to a defect can lead to simultaneous appearance of discrete levels and deformation of the quasi-continuous spectrum due to the localized states that split off from the boundary of the spectrum. The deformation can be described by the shift function [20].

The change in heat capacity due to one gap level is described by the expression:

$$\Delta C_V(T) = \sum_i \left( F(\varepsilon_{i\,g}, T) - F(\varepsilon_{i\,min}, T) - \int_{\varepsilon_{i\,min}}^{\varepsilon_{i\,max}} F(\varepsilon, T) \frac{\partial \xi_i(\varepsilon)}{\partial \varepsilon} d\varepsilon \right)$$

$$F(\varepsilon, T) = k_B \left( \frac{\hbar \sqrt{\varepsilon}}{2 k_B T} \right)^2 \text{sh}^{-2} \left( \frac{\hbar \sqrt{\varepsilon}}{2 k_B T} \right),$$

(6)

The main change in the low-temperature heat capacity (Fig. 7) is determined by the direct contribution of the gap energy levels. We consider the chains of atoms with a low concentration of defects. In this case, we can consider the problem in a linear approximation with respect to the concentration $p$. The defect in question has the greatest overall effect on the change in the ratio of the lengths of parts of the heat capacity temperature dependence curve in the case of a compressed chain ($\beta_\tau < 0, \beta_l > 0$).

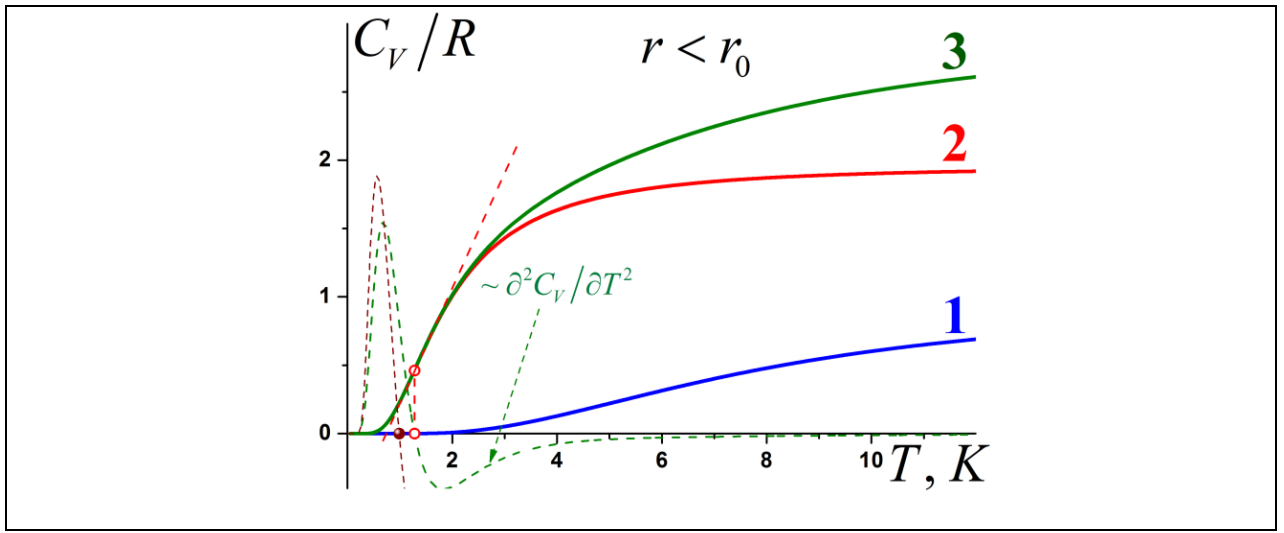

**Fig 7:** Phonon heat capacity of a compressed chain ($r < r_0$). Curve 1 is the contribution to the heat capacity of the longitudinal vibrations; curve 2 – the contribution of the transverse vibrations; curve 3 is the total heat capacity. Dashed line the value proportional to $\dfrac{\partial^2 \tilde{N}_V}{\partial T^2}$ for the chain without defects. Dash-dotted line is the tangent to $\tilde{C}_V$ at the inflection point. Dotted line is the value proportional to $\dfrac{\partial^2 \tilde{C}_V(T)}{\partial T^2}$.

The shape of the curves is also influenced by the term (6), which contains the shift function $\xi(\varepsilon)$. The expression for the shift function of the linear chains in terms of the Jacobian matrix method was obtained in [21]. In this article, we give the expressions for the shift functions generated by the defect in question. In the subspace $H^{(-)}$, the shift function has the form:

$$\xi_i^{(-)}(\varepsilon) = \frac{1}{\pi} \operatorname{arcctg}\left( \frac{2b_i - \lambda_i^{(-)}}{\lambda_i^{(-)}} \sqrt{\frac{\varepsilon - \varepsilon_{i\min}}{\varepsilon_{i\max} - \varepsilon}} \right); \quad \beta_i > 0, \tag{7}$$

$$\xi_i^{(-)}(\varepsilon) = \frac{1}{\pi} \operatorname{arcctg}\left( \frac{2|b_i| + \lambda_i^{(-)}}{\lambda_i^{(-)}} \sqrt{\frac{\varepsilon_{i\max} - \varepsilon}{\varepsilon - \varepsilon_{i\min}}} \right); \quad \beta_i < 0. \tag{8}$$

In the subspace $H^{(+)}$, the following expressions are obtained for the shift function:

$$\xi_i^{(+)}(\varepsilon) = \frac{1}{\pi} \operatorname{arcctg}\left( \frac{2b_i + \lambda_i^{(+)}}{\lambda_i^{(+)}} \sqrt{\frac{\varepsilon_{i\max} - \varepsilon}{\varepsilon - \varepsilon_{i\min}}} \right); \quad \beta_i > 0, \tag{9}$$

$$\xi_i^{(+)}(\varepsilon) = \frac{1}{\pi} \operatorname{arcctg}\left( \frac{2|b_i| - \lambda_i^{(+)}}{\lambda_i^{(+)}} \sqrt{\frac{\varepsilon - \varepsilon_{i\min}}{\varepsilon_{i0} - \varepsilon}} \right); \quad \beta_i < 0. \tag{10}$$

The full shift function can be written as follows:

$$\xi(\varepsilon) = \xi_l^{(-)}(\varepsilon) + \xi_l^{(+)}(\varepsilon) + 2\left[\xi_\tau^{(-)}(\varepsilon) + \xi_\tau^{(+)}(\varepsilon)\right]. \tag{11}$$

The density of states of a chain with defects in the linear approximation with respect to the concentration $p$ is:

$$g_{id}(\varepsilon) = g_i(\varepsilon) - p\frac{d\xi_i(\varepsilon)}{d\varepsilon}, \tag{12}$$

where $g(\varepsilon)$ is the density of states in the absence of defects. The case of a compressed chain is of greatest interest in the problem of the shift of the linear-like part of the heat capacity temperature dependence. In the case of a small perturbation ($\lambda_i^{(\pm)} \ll b_i$) in a compressed chain, the change of its vibrational density of states is:

$$\Delta g(\varepsilon) = g_d(\varepsilon) - g(\varepsilon) = p\left(\frac{\lambda_l^{(-)}}{\varepsilon - \varepsilon_{l0}} - \frac{\lambda_l^{(+)}}{\varepsilon_{l\max} - \varepsilon}\right) g_l(\varepsilon) + 2p\left(\frac{\lambda_\tau^{(+)}}{\varepsilon - \varepsilon_{\tau\min}} - \frac{\lambda_\tau^{(-)}}{\varepsilon_{\tau\max} - \varepsilon}\right) g_\tau(\varepsilon), \quad (13)$$

where

$$g_i(\varepsilon) = \frac{1}{2\pi} \operatorname{Im}\left\{G_i^{(-)}(\varepsilon) + G_i^{(+)}(\varepsilon)\right\}, \quad i = l, \tau. \quad (14)$$

It is seen that due to the presence of defects, the vibrational density of states changes mainly at the points of singularity. It directly changes the type of the singularities.

In the case of not a small perturbation created by the defect, the change in the vibrational densities of a stressed chain in the vicinity of the singularity points for each of the vibration branches is proportional to the vibrational density of the unperturbed chain for each of the vibration branches.

At $\Delta_i > 0$:

$$\lim_{\varepsilon \to \varepsilon_{i\max}} \Delta g_i(\varepsilon) = p \cdot \left(\frac{\lambda_i^{(-)}}{2b_i - \lambda_i^{(-)}} - \frac{2b_i + \lambda_i^{(+)}}{\lambda_i^{(+)}}\right) \frac{g_i(\varepsilon_{i\max})}{2};$$

$$\lim_{\varepsilon \to \varepsilon_{i\min}} \Delta g_i(\varepsilon) = p \cdot \left(\frac{2b_i - \lambda_i^{(-)}}{\lambda_i^{(-)}} - \frac{\lambda_i^{(+)}}{2b_i + \lambda_i^{(+)}}\right) \frac{g_i(\varepsilon_{i\min})}{2};$$

(15)

at $\Delta_i < 0$:

$$\lim_{\varepsilon \to \varepsilon_{i\max}} \Delta g_i(\varepsilon) = p \cdot \left(\frac{\lambda_i^{(+)}}{2|b_i| - \lambda_i^{(+)}} - \frac{2|b_i| + \lambda_i^{(-)}}{\lambda_i^{(-)}}\right) \frac{g_i(\varepsilon_{i\max})}{2};$$

$$\lim_{\varepsilon \to \varepsilon_{i\min}} \Delta g_i(\varepsilon) = p \cdot \left(\frac{2|b_i| - \lambda_i^{(+)}}{\lambda_i^{(+)}} - \frac{\lambda_i^{(-)}}{2|b_i^{(-)}| + \lambda_i^{(-)}}\right) \frac{g_i(\varepsilon_{i\min})}{2}.$$

(16)

Figure 7 shows the shift of the linear part of the temperature dependence of the heat capacity of a compressed chain with defects $\tilde{C}_v(T) = C_v(T) + p\Delta C_v(T)$ toward low temperature due to the direct contribution of defects at $p = 0.05$, $\delta_\tau = -0.61$, $\eta_\tau = -0.65$.

## Conclusions

The stability of structure of the adsorbed linear chains is conditioned mainly by their interaction with the substrate. Thus the interatomic distance in the chains differs from the equilibrium distance of the potential of the interatomic interaction between atoms of the gas forming the chain, and the chain itself is stressed. If the distance between atoms in the chain is smaller than the equilibrium one (the chain is "compressed"), then the quasi-continuous spectrum band of its transverse vibrations shifts to low frequencies. In this case, the linear part of the temperature dependence of the heat capacity shifts to lower temperatures.

Both the stressed state of the adsorbed chain and defects of nanotubes lead to the appearance of isolated defects in the chain, which are local changes in the interatomic distance. For transverse vibrations of the "compressed" chains it leads to a non-threshold formation of discrete vibrational levels with frequencies lying below the quasi-continuous spectrum band, which results in further extension of the linear part on the temperature dependence of the heat capacity to an even lower temperatures.

**References:**

1. L.D. Landau, *JETP* **7**, 627 (1937).
2. J.V. Pearce, M.A. Adams, O.E. Vilches, M.R. Jonson, and H.R. Glyde, *Phys. Rev. Lett*. **95**, 185302 (2005).
3. M.I. Bagatskii, M.S. Barabashko, A.V. Dolbin, and V.V. Sumarokov, *Low Temp. Phys*. **38**, 523 (2012) [*Fizika Nizkikh Temperatur* **38**, 667–673 (2012)].
4. M.I. Bagatskii, M.S. Barabashko, V.V. Sumarokov, *Low Temp. Phys*. **39**, 441 (2013) [*Fizika Nizkikh Temperatur* **39** 568 (2013)].
5. M.I. Bagatskii, V.G. Manzhelii, V.V. Sumarokov, and M.S. Barabashko, *Low Temp. Phys*. **39**, 618 (2013) [*Fizika Nizkikh Temperatur* **39**, 801 (2013)]
6. M.I. Bagatskii, M.S. Barabashko, V.V. Sumarokov, *JETP Lett*. **99**, 461 (2014) [*Письма в ЖЭТФ*, 99, 532].
7. M.I. Bagatskii, M.S. Barabashko, V.V. Sumarokov, A. Jeżowski, P Stachowiak, *JLTP*, **187**, 113 (2017).
8. Šiber, *Phys. Rev. B* **66**, 235414 (2002).
9. M. Aichinger, S. Kilić., E. Krotscheck, and L. Vranješ, *Phys. Rev.* **B 70**, 155412 (2004).

10. E.V. Manzhelii, S.B. Feodosyev, I.A. Gospodarev, E.S. Syrkin and K.A. Minakova, *Low. Phys.*, **41**, 557 (2015) [*Fizika Nizkikh Temperatur* **41**, 718 (2015)].
11. E.V. Manzhelii, *JLTP*, **187**, 105 (2017).
12. . E.V. Manzhelii, *Low Temp. Phys*. **29**, 333 (2003) *[Fizika Nizkikh Temperatur* **29**, 801 (2003)].
13. S.B. Feodosyev, I.A. Gospodarev, V.O. Kruglov, E.V. Manzhelii, *JLTP*, **139**, 651, (2005).
14. K.A. Chishko and E.S. Sokolova, *Low. Temp. Phys*. 42, 85 (2016)[*Fizika Nizkikh Temperatur* **42**, 116 (2016)].
15. V. I. Peresada, V. N. Afanas'ev, and V. S. Borovikov, *Sov. J. Low Temp. Phys*. **1**, 227 (1975)
16. M.A. Mamalui, E.S. Syrkin, S.B. Feodosyev, *Low Temp. Phys*. **25**, 732 (1999) [*Fizika Nizkikh Temperatur*, **25**, 976 (1999)].
17. M.A. Mamalui, E.S. Syrkin, S.B. Feodosyev, *Low Temp. Phys*. **25**, 55 (1999) [*Fizika Nizkikh Temperatur* **25**, 72 (1999)].
18. A.V. Kotlyar, and, S.B. Feodosyev, *Low Temp. Phys*. **32**, 256 (2006) [*Fizika Nizkikh Temperatur*, **32**, 343 (2006)].
19. I. A. Gospodarev, A. V. Grishaev, E. S. Syrkin, and S. B. Feodos'ev, *Physics of Solid State*, **42**, 2217 (2000) [*ФТТ* **42,** 2153 (2000)].
20. I.M. Lifshits *Usp. Mat. Nauk* **7**, 171 (1952) [in Russian].
21. E.S. Syrkin, S.B. Feodosyev, *Fizika Nizkikh Temperatur*, **20**, 586 (1994) [in Russian].